\documentclass[a4paper,twocolumn,superscriptaddress,11pt,accepted=2019-03-01]{quantumarticle}
\pdfoutput=1
\usepackage[utf8]{inputenc}
\usepackage[english]{babel}
\usepackage[T1]{fontenc}
\usepackage{amsmath}
\usepackage{hyperref}
\usepackage{txfonts}
\usepackage{amssymb}
\usepackage{amsmath}

\usepackage{CJK}
\usepackage{epsfig}
\usepackage{url}
\usepackage{subfigure}
\usepackage{multirow}
\usepackage{bm}
\usepackage{tikz}
\usepackage{lipsum}
\usepackage{hyperref}
\hypersetup{  colorlinks=true,           linkcolor=blue,              citecolor=red,            urlcolor=blue  }

\renewcommand\refname{References and Notes}

\begin{document}

\title{Emergence of Network Bifurcation Triggered by Entanglement}

\author{Xi Yong}
\affiliation{State Key Laboratory of Computer Science, Institute of Software, Chinese Academy of Sciences, P. R. China}
\affiliation{University of Chinese Academy of Sciences, P. R. China}
\affiliation{Water Information Center, Ministry of Water Resources, Beijing 100053, P. R. China}

\author{Man-Hong Yung}
\email{yung@sustc.edu.cn}
\affiliation{Institute for Quantum Science and Engineering and Department of Physics, Southern University of Science and Technology, Shenzhen 518055, P. R. China}
\affiliation{Shenzhen Key Laboratory of Quantum Science and Engineering, Shenzhen 518055, P. R. China}
\affiliation{Center for Quantum Information, Institute for Interdisciplinary Information Sciences, Tsinghua University, P. R. China}
\affiliation{Central Research Institute, Huawei Technologies, Shenzhen 518129, P. R. China}

\author{Xue-Ke Song}
\email{songxk@sustc.edu.cn}
\affiliation{Institute for Quantum Science and Engineering and Department of Physics, Southern University of Science and Technology, Shenzhen 518055, P. R. China}
\affiliation{Shenzhen Key Laboratory of Quantum Science and Engineering, Shenzhen 518055, P. R. China}
\affiliation{Department of Physics, Southeast University, Nanjing 211189, P. R. China}

\author{Xun Gao}
\affiliation{Center for Quantum Information, Institute for Interdisciplinary Information Sciences, Tsinghua University, P. R. China}

\author{Angsheng Li}
\email{angsheng@ios.ac.cn}
\affiliation{State Key Laboratory of Computer Science, Institute of Software, Chinese Academy of Sciences, P. R. China}

\maketitle

\begin{abstract}
In many non-linear systems, such as plasma oscillation, boson condensation,  chemical reaction, and even predatory-prey oscillation, the coarse-grained dynamics are governed by an equation containing anti-symmetric transitions, known as the anti-symmetric Lotka-Volterra (ALV) equations. In this work, we prove the existence of a novel bifurcation mechanism for the ALV equations, where the equilibrium state can be drastically changed by flipping the stability of a pair of fixed points. As an application, we focus on the implications of the bifurcation mechanism for evolutionary networks; we found that the bifurcation point can be determined quantitatively by the microscopic quantum entanglement. The equilibrium state can be critically changed from one type of global demographic condensation to another state that supports global cooperation for homogeneous networks. In other words, our results indicate that there exist a class of many-body systems where the macroscopic properties are invariant with a certain amount of microscopic entanglement, but they can be changed abruptly once the entanglement exceeds a critical value. Furthermore, we provide numerical evidence  showing that the emergence of bifurcation is robust against the change of the network topologies, and the critical values are in good agreement with our theoretical prediction. These results show that the bifurcation mechanism could be ubiquitous in many physical systems, in addition to evolutionary networks.
\end{abstract}

\section{Introduction}
 The non-linear dynamics of ecological systems where multiple species interact is commonly described by the Lotka-Volterra equations~\cite{Yorke1973} (also known as the predator-prey equations), which contains a set of coupled first-order, non-linear, differential equations. It turns out that the Lotka-Volterra equations finds a broad range of applications beyond ecological systems, and has created a profound impact on physical sciences~\cite{Nutku1990,Kerner1990,Matsuda1992,Malcai2002,Mobilia2007}. In this work, we are interested in a specific class of the Lotka-Volterra equations, where the Malthusian growth/decay term is negligible and the transition coefficients are all anti-symmetric. The resulting equations (see Eq.~(\ref{ALV_eq})) are referred to as anti-symmetric Lotka-Volterra (ALV) equations~\cite{Knebel2015,Reichenbach2006,Vorberg2013,Zakharov1974,DiCera1988}. The ALV equations describe a variety of interesting processes, including predatory-prey oscillations in population biology~\cite{Reichenbach2006}, boson condensation far from equilibrium~\cite{Vorberg2013}, plasma oscillation~\cite{Zakharov1974}, kinetics of chemical reactions~\cite{DiCera1988}, etc.

\begin{figure}[t]
    \centering
    \includegraphics[width=0.7\columnwidth]{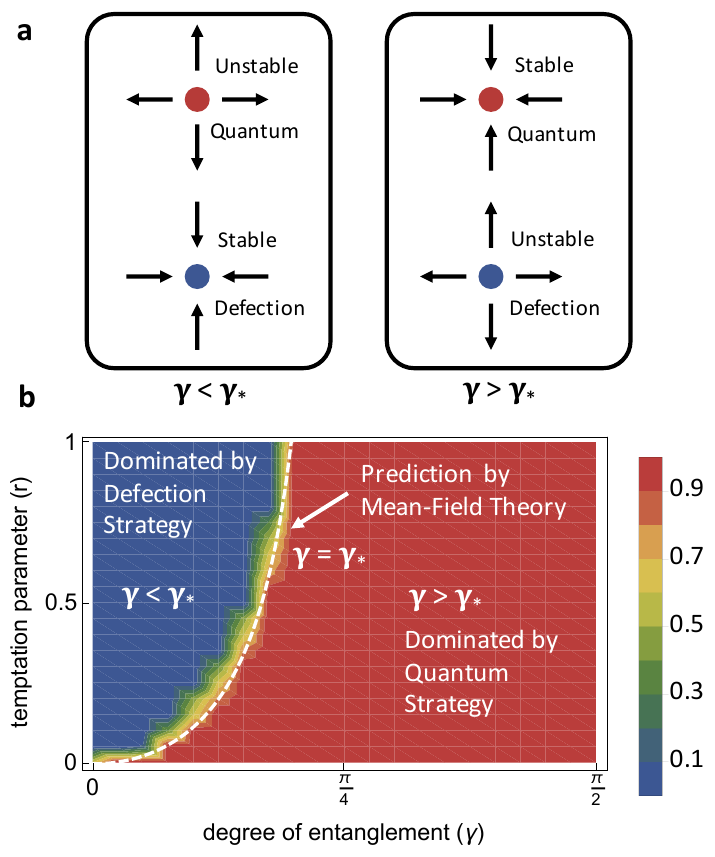}
    \caption{Fixed-point stability and phase diagram obtained by numerical simulation of network of square lattice (see Eq.~(\ref{dynamical_eom_mf})). (a) The stability of two strategies, Defection and Quantum depends on the value of the parameter $\gamma$. (b) The color bar represents the value of the density of Quantum strategy. When $\gamma<\gamma_*$, the network is dominated by the Defection strategy. When $\gamma > \gamma_*$, the network is dominated by the Quantum strategy. The white dotted line comes from the prediction on the phase boundary by our mean-field theory (see Eq.~(\ref{eq:phase_boundary})).}
    \label{fig:phase}
\end{figure}

Here, we prove that there exists a novel bifurcation mechanism predicted by the ALV equations, formed by flipping the stability of a pair of fixed points (see Fig.~\ref{fig:phase}). Specifically, we study its implications on evolutionary networks as a concrete application; we found that the bifurcation mechanism can be triggered by varying the amount of quantum entanglement~\cite{Horodecki2009a} in the microscopic interaction in the evolutionary networks, which represents a quantum solution to the problem of promoting global cooperation in homogeneous networks.

%The theory of evolutionary network models a many-body system of interacting nodes (or decision-making agents) in a network. The microscopic interaction is realized as a form a ``game" played between the agents, whose primary goal is to increase their individual utilities or payoffs. Models in evolutionary network are typically applied to studying the dynamics of repeated games played by rational agents, where the evolutionarily stable strategy determines the stability of the equilibrium under fluctuations, and provides a framework for understanding the emergence of the cooperative behaviors in nature and society, which plays an essential role in biosphere and human society.

Over the past few decades, evolutionary network theory~\cite{NM1992} has become the main paradigm of connecting non-equilibrium statistical physics to many other scientific areas including biology, psychology, economics, and behavioral sciences (see e.g. Ref.~\cite{Szabo2007} for a review).
%It provides a set of analytic tools to explain and predict the outcomes of networks of competing agents, and leads to the development of new scientific disciplines such as socio- and econo-physics.
Real-world networks are realized in systems characterized by graphs in which the vertices (or nodes) represent players, and the edges represent the games played. One of the major goals of evolutionary network theory is to understand the main factors that affect the emergence of {\it global cooperation}. It is known that network topology is crucial for the long-time behaviors of evolutionary networks~\cite{NM1992,HS2005,Szabo2007,Gomez-Gardenes2007,RCS2009,yukalov2018}.
%In particular, it was found~\cite{NM1992} that spatial reciprocity (or spatial structure) plays a key role in promoting cooperation.
Particularly, it was recognized that {\it heterogeneity} (e.g. existence of large hubs) of networks plays a role in the emergence of cooperations of evolutionary games in networks~\cite{F1974}.
%This feature was found in scale-free network~\cite{Ba1999, SP2005, SP2006} and in the public goods games~\cite{SSP2008}.
%Very often, cooperations emerge around the largest hub, as evident in many different scenarios~\cite{ST1998, SP2005, PS2005, SPL2006, W2006, SPL2007, GGCFM2007,SSP2008}.
However, for many {\it homogeneous} networks, which is of interest in this work, it remains a major challenge to achieve global cooperation~\cite{Szabo1998,Wu2006, Nowak2006, Vukov2008}. A central open question in this area is how one can increase the global payoffs of evolutionary games in homogeneous networks.

The existence of the bifurcation mechanism implies that for a quantum model of evolutionary network, by varying the amount of quantum entanglement~\cite{Horodecki2009a} in the underlying interaction, one can unambiguously achieve the goal of enhancing global cooperation for homogenous networks, avoiding the need of including large hubs.

Moreover, our result implies that the global behaviors of the network can be insensitive to a certain amount of entanglement involved in the microscopic interaction, but they can be abruptly changed, when the entanglement in the microscopic interaction exceeds a critical value. Away from the critical point, we found that the equilibrium state forms a demographic condensations with very weak fluctuations, which resembles Bose-Einstein condensation in statistical physics.

In addition, we show that the non-linear dynamics of the anti-symmetric Lotka-Volterra equation can be constructed by a mean-field theory on the evolutionary networks, which describes the process as a dynamical phase transition in the network; depending on the amount of quantum entanglement involved, the equilibrium state can be driven from an unordered phase to one out of the two condensed phases.

The origin of the bifurcation is found to be related to a critical scenario where a pair of distinct fixed points exchange their roles from being stable to unstable and vice versa. In our model of evolutionary network, this scenario is possible only if quantum entanglement are provided. Finally, our numerical simulations indicate that the emergence of bifurcation is robust against the change of the network topologies, and the critical values are in good agreement with the theoretical prediction from our mean-field analysis. These results suggest that the bifurcation mechanism could exist in systems other than evolutionary networks; our analysis serves as a guide for finding bifurcation points in other systems governed by the ALV equation.

\section{Anti-symmetric Lotka-Volterra equation}
 The anti-symmetric Lotka-Volterra equation~\cite{Zakharov1974,DiCera1988,Reichenbach2006,Vorberg2013,Knebel2015} is given by the following expression,
\begin{equation}\label{ALV_eq}
  \frac{d}{{dt}} \ {\rho_i} = {\rho_i} \sum\limits_{j \ne i} {{A_{ij}} \ {\rho_j}} \ ,
\end{equation}
where $\rho_i \geq 0$ is the value of some quantity of interest (e.g. population of a species $i$). The elements of the anti-symmetric matrix are given by
\begin{equation}
{A_{ij}} = {w_{ij}} - {w_{ji}} = - A_{ji} \ ,
\end{equation}
{where $w_{ij}$ is the transition rate describing an agent changes his/her strategy from $i$ to $j$ for a given evolutionary network}.

It is sufficient to demonstrate our results in the situations involving only three species, i.e., $i=1,2,3$. We shall show that, except for some singular points, in general, there are exactly (i) one stable, (ii) one unstable, and (iii) one saddle fixed points. {More specifically, in the steady-state solutions of the ALV equation in the long time limit, i.e., $ {\rho _1^\infty } \equiv \rho_1(t {\to} \infty)$, ${\rho _2^\infty } \equiv \rho_2(t {\to} \infty)$, and ${\rho _3^\infty } \equiv \rho_3(t {\to} \infty)$, we have $\tfrac{d}{{dt}}{\rho _1} = \tfrac{d}{{dt}}{\rho _2} = \tfrac{d}{{dt}}{\rho _3} = 0$. In the following, we will see that there exist three fixed points, namely,}
{\begin{equation}
\left[ {\begin{array}{*{20}{c}}
  {\rho _1^\infty } \\
  {\rho _2^\infty } \\
  {\rho _3^\infty }
\end{array}} \right] = \left\{ \ { \ \left[ {\begin{array}{*{20}{c}}
  1 \\
  0 \\
  0
\end{array}} \right],\left[ {\begin{array}{*{20}{c}}
  0 \\
  1 \\
  0
\end{array}} \right],\left[ {\begin{array}{*{20}{c}}
  0 \\
  0 \\
  1
\end{array}} \right]} \ \right\} \ .
\end{equation}}

{At first, we have the following relations from the Eq.~(\ref{ALV_eq}):}
{\begin{align}
  \rho _1^\infty \ \rho _2^\infty \ {A _{12}} + \rho _1^\infty \ \rho _3^\infty \ {A_{13}} =  & \ 0 \ , \label{rho_CD_CQ} \\
   - \rho _2^\infty \ \rho _3^\infty \ {A _{23}} + \rho _2^\infty \ \rho _1^\infty \ {A _{21}} =  & \ 0 \ , \label{rho_DQ_DC} \\
  \rho _3^\infty \ \rho _1^\infty \ {A _{31}} + \rho _3^\infty \ \rho _2^\infty \ {A _{32}} =  & \ 0 \label{rho_QC_QD} \ ,
\end{align}}
{which subject to the constraint, }
{\begin{equation} \label{normalization}
\rho _1^\infty  + \rho _2^\infty  + \rho _3^\infty  = 1 .
\end{equation}}

{Since the transition rate $w_{ij}$ being non-zero implies that $A_{ij}$ is also non-zero. Then, Eq.~(\ref{rho_CD_CQ}) means that $\rho _1^\infty \rho _2^\infty  = \rho _1^\infty \rho _3^\infty  = 0$, which shows that either (i) $\rho _1^\infty =0$ or (ii) a fixed point at $\rho _1^\infty =1 $ and $\rho _2^\infty  = \rho _3^\infty  = 0$. The case (i) implies either $\rho _2^\infty =0 $ or $\rho _3^\infty =0 $ from Eqs.~(\ref{rho_DQ_DC}), (\ref{rho_QC_QD}), and (\ref{normalization}), which means that ${[\rho _1^\infty ,\rho _2^\infty ,\rho _3^\infty ]^T} = \{ {{{[0,1,0]}^T},{{[0,0,1]}^T}} \}$ are  fixed points as well. The theory can be extended to higher number of families by doing the same calculations.}

%We first locate the fixed points of the ALV equation: by imposing the normalization condition,
%\begin{equation}
%\sum\limits_i \ {{\rho _i}}  = 1 \ ,
%\end{equation}
%and  setting all $d \rho_i / dt=0$ in the long-time limit, a set of fixed points $\bm {\rho _{{\text{fix}}}} = \{ \ {\rho _i^\infty} \ \} $, where $\rho _i^\infty  \equiv {\rho _i}\left( {t \to \infty } \right)$, can be located by choosing a final state dominated by one of the species, i.e, $\rho _i^\infty  = 1$ and $\rho _{j \ne i}^\infty=0$ for all $j \ne i$.
%
%\red{Since the transition rate $w_{ij} \ne 0$ being non-zero implies that $A_{ij} \ne 0 $ is also non-zero. Then the Eq.~(\ref{ALV_eq}) implies that
%\begin{equation}
%\rho_i^\infty\rho_j^\infty=0
%\end{equation}
%for all $j \ne i$, which means that either (i) $\rho_i^\infty=0$ or (ii) a fixed point $\rho_i^\infty=1$ and $\rho_j^\infty=0$ ($j \ne i$) due to the normalization condition. The case (i) implies $\rho_j^\infty=1$ and  $\rho_k^\infty=0$ with $k \ne j$ from the ALV equation, and the case (ii) means that $\rho_j^\infty=0$ for all $j \ne i$. This implies that the fixed points are located at the points $\rho _i^\infty  = 1$ and $\rho _{j \ne i}^\infty=0$ for all $j \ne i$.}

\section{Fixed-point stability of ALV equation}
 Around one of the fixed points, the linearized differential equations is given by
%$  d {\rho _i}/dt = {\bm \nabla} ( {{\rho _i}\sum\nolimits_{j \ne i} {{A_{ij}} \ {{\rho} _j}} } ) \cdot \left( {{\bm \rho}  - {{\bm \rho}_{{\text{fix}}}}} \right)$,
\begin{equation}
  \frac{d}{{dt}}{\rho _i} = {\bm \nabla} ( {{\rho _i}\sum\limits_{j \ne i} {{A_{ij}} \ {{\rho} _j}} } ) \cdot \left( {{\bm \rho}  - {{\bm \rho}_{{\text{fix}}}}} \right) \ ,
\end{equation}
where the gradient, ${\bm \nabla}  \equiv (\partial /\partial {\rho _1},\partial /\partial {\rho _2},...)$, is evaluated at the corresponding fixed point. Furthermore, {the corresponding Jacobian matrices, labeled by $ \{ J_{1}, J_{2}, J_{3} \}$ for the three fixed points, namely ${\left[ {1,0,0} \right]^T}$, ${\left[ {0,1,0} \right]^T}$, and ${\left[ {0,0,1} \right]^T}$, respectively, are listed as follows:}
\begin{equation} \label{jocabian}
\left \{ \left[ {\begin{array}{*{20}{c}}
  0&{{A_{12}}}&{{A_{13}}} \\
  0&{{A_{21}}}&0 \\
  0&0&{{A_{31}}}
\end{array}} \right], \ \left[ {\begin{array}{*{20}{c}}
  {{A_{12}}}&0&0 \\
  {{A_{21}}}&0&{{A_{23}}} \\
  0&0&{{A_{32}}}
\end{array}} \right],  \ \left[ {\begin{array}{*{20}{c}}
  {{A_{13}}}&0&0 \\
  0&{{A_{23}}}&0 \\
  {{A_{31}}}&{{A_{32}}}&0
\end{array}} \right] \right \} .
\end{equation}
The eigenvalue spectra $\lambda \left( {{J_i}} \right)$ of the matrices are found to be
%$\lambda \left( {{J_i}} \right) \ \in \ \{ \ {0, \ {A_{ji}}, \ {A_{ki}}} \ \} $
{\begin{equation}
\lambda \left( {{J_i}} \right) \ \in \ \{ \ {0, \ {A_{ji}}, \ {A_{ki}}} \ \}
\end{equation}}
for distinct values of $j \ne k \ne i$. {Recall that a fixed point in ALV equation is stable (unstable) when the remaining two eigenvalues are both negative (positive); otherwise it is a saddle fixed point. The zero-eigenvalue exists for all Jacobian matrices, which results from the constraint of the normalization condition, i.e., $\rho _1 + \rho _2 +\rho _3 = 1$.}

Now, suppose each node (or player) $i$ is associated with a value, $P_i$ (e.g. energy or payoff), where the transition rates $w_{ij}$ are a monotonic function of the difference between the values of $P$'s, i.e., ${w_{ij}} = w( {{P_i} - {P_j}} )$. A common example for the transition rate is given by the Fermi function~\cite{Szabo2007},
\begin{equation}\label{Fermi_function}
{w_{ij}} = 1/{(1 + {e^{ - ({P_j} - {P_i})/T}})} \ ,
\end{equation}
where $T$ is a parameter for controlling the transition rates. We note that if $P_j - P_i>0$, then ${w_{ij}} > {w_{ji}}$, which implies that $A_{ij}$ ($A_{ji}$) is positive (negative), i.e., $A_{ij} > 0 $ and $A_{ji} < 0 $.

In general, the values of $P$'s are non-degenerate in equilibrium, except for some singular points, which means that there exists an order, e.g., ${P_3} > {P_2} > {P_1}$. In this case,
{\begin{enumerate}
  \item[(i)] the fixed point ${\left[ {0,0,1} \right]^T}$ is stable, as $A_{13}<0$ and $A_{23}>0$;
  \item[(ii)] ${\left[ {0,1,0} \right]^T}$ is a saddle point, as $A_{12} > 0$ but $A_{32} < 0$;
  \item[(iii)] ${\left[ {1,0,0} \right]^T}$ is unstable, as $A_{21} < 0$ and $A_{31} <0$.
\end{enumerate}}
Similar results can be obtained by any permutation of the values of $P$'s.

\medskip
{\noindent \bf Guiding principles for the bifurcation mechanism---} Let us focus on the following scenario: suppose there exist a parameter $\gamma$ (to be identified as the amount of entanglement for evolutionary networks) such that when it is smaller than a critical value~$\gamma_{*}$, i.e., $\gamma < \gamma_{*}$, we have one particular order of the $P$ values, e.g., $P_2 > P_3 > P_1$. Based on our analysis above, it means that the long-time population of the anti-symmetric Lotka-Volterra equation should be dominated by the state $[0,1,0]^T$. Furthermore, suppose, whenever we increase the value of $\gamma_{*}$ to cross the critical point, i.e.,  $\gamma > \gamma_{*}$, we have a different order, e.g., $P_3 > P_2 > P_1$; consequently, we shall be able to observe that the equilibrium state {\it abruptly} changes to another state $[0,0,1]^T$, even though the change the parameter $\gamma$ is smooth.

Instead of an abstract formalism, in the following, we provide a concrete instance on how such a bifurcation mechanism can indeed occur in the context of evolutionary networks, where we prove that the parameter~$\gamma$ can be characterized by the amount of entanglement in the microscopic interactions between the nodes in the network. In principle, for other dynamical systems governed by the ALV equation, if the corresponding transition rates contain an adjustable parameter that can be varied as described in our model, a bifurcation point can also be located in the same manner, with or without entanglement.

\section{Evolutionary network as application}
\begin{figure}[t]
    \centering
    \includegraphics[width=1\columnwidth]{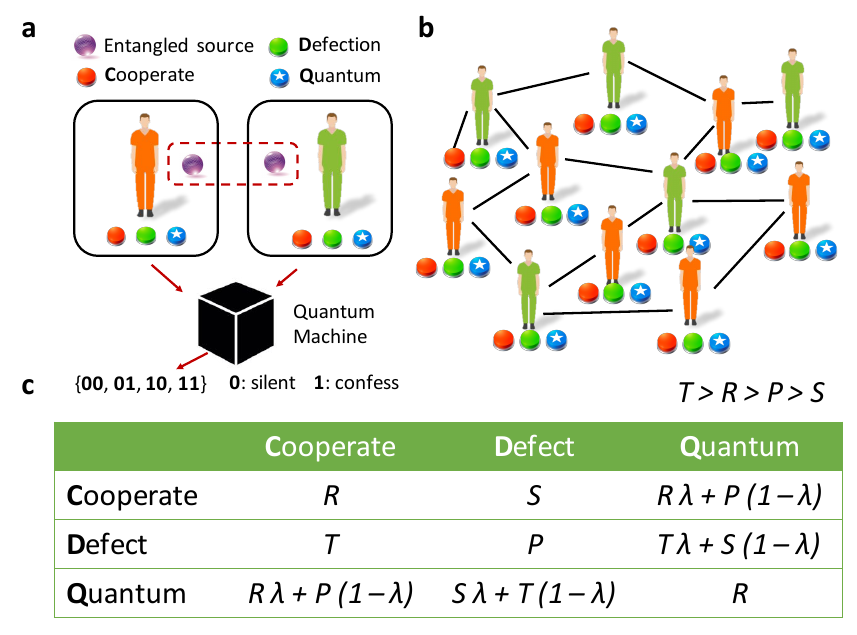}
    \caption{Microscopic interaction in the evolutionary network. (a) The players are given an entangled pair and three choices \{Cooperate, Defect, Quantum\} for each game; their choice will be sent to a quantum machine to probabilistically make the final decisions for them. Mention that the entangled state
doesn't let the two parties communicate. (b) The games are played by all members in a network. (c) The payoff table depends on the amount of entanglement with $\lambda=\cos^2\gamma$ (see Eq.~(\ref{eq:PI00011011})) .}
    \label{fig:prisoner}
\end{figure}

The physical model of the evolutionary network is defined as follows~\cite{Szabo2007,HS2005}. (i) There are $N \gg 1$ rational and identical agents located on the sites in a network; (ii) They interact (formalized as a two-player game) repeatedly with their neighbors to gain/lose an income. (iii) After each game, the agents are allowed to change their strategies in order to increase their utility. In particular, they tend to learn their neighbor's strategies that has generated a higher income. Here a game (see Fig.~\ref{fig:prisoner} (b)) is an abstract formulation of an interaction among $N$ players that have potential conflicting interests. The updating probability of a node $i$ to adopt the strategy ${\cal S}_j$ of a reference node $j$, is determined by a transition probability, $w_{ij} \equiv w( {{{\cal S}_i} \to {{\cal S}_j}} )$, which is a function of the difference of the payoffs $P_i - P_j$, and is defined to be identical to the Fermi function (see Eq.~(\ref{Fermi_function})).

Since the local games are played probabilistically, the macroscopic configurations form a probabilistic distribution; we define $Q({\bf n},t)$ to be the probability for the configuration $\bf n$ to appear at time $t$. The rate of change of $Q({\bf n},t)$ is determined by the inflow into and outflow from the configuration $\bf n$, which means that  the dynamics of the evolutionary network can be described by a master equation,
\begin{equation}\label{master_eq}
  \frac{d}{{dt}}Q\left( {{\bf n},t} \right) = \sum\limits_{{\bf n}'} {\left[ {Q\left( {{\bf n}',t} \right){W_{{\bf n}' \to {\bf n}}} - Q\left( {{\bf n},t} \right){W_{{\bf n} \to {\bf n}'}}} \right]} \ ,
\end{equation}
where ${W_{{\bf{n}} \to {\bf{n}}'}}$ is the configurational transition rate from the configuration $\bf n$ to $\bf n '$ and plays the main role for the dynamics.

The configurational transition rate ${W_{{\bf{n}} \to {\bf{n}}'}}$ comes from the microscopic transition rate $W^\mu_{x \to y}{\left( {\bf{n}} \right)}$, which describes an agent changes his strategy from $\mathcal S_x$ to $\mathcal S_y$ for a given macroscopic configuration $\bf n$. The configurational transition rate is proportional to the number $n_x$ of agents with $\mathcal S_x$. Explicitly, we can write
%${W_{{\bf{n}} \to {\bf{n}}'}} = \sum\nolimits_{x,y} {{n_x}\;W^{\mu}_{x \to y} {\left( {\bf{n}} \right)} \ {\delta _{{\bf{n}}',{{\bf{n}}_{x \to y}}}}}$,
\begin{equation}\label{Wnnp_nx_wxy_nxy}
{W_{{\bf{n}} \to {\bf{n}}'}} = \sum\limits_{x,y} {{n_x}\;W^{\mu}_{x \to y} {\left( {\bf{n}} \right)} \ {\delta _{{\bf{n}}',{{\bf{n}}_{x \to y}}}}} \ ,
\end{equation}
where $x$ and $y \in \left\{ {a,b,c, \cdots } \right\}$ and the vector ${{\bf{n}}_{x \to y}} = [\cdots, n_x {-}1, \cdots, n_y {+}1, \cdots ]^T$ labels the configuration that is obtained from $\bf n$ by changing $n_x \to n_x -1$ and $n_y \to n_y +1$.

\section{ALV equation as a mean-field theory}
Based on the discussion above, we can obtain a dynamical equation,
\begin{equation}
\frac{d\left\langle {{n_x}}\right\rangle}{dt} = \sum\limits_{{\bf{n}},{\bf{n}}{'}} {({n'_x} - {n_x}){W_{{\bf{n}} \to {\bf{n}}'}} Q\left( {{\bf{n}},t} \right)} \ ,
\end{equation}
for the average occupation of each strategy, where $\left\langle {{n_x}} \right\rangle  \equiv \sum\nolimits_{\bf{n}} {{n_x} \ Q\left( {{\bf{n}},t} \right)} $ (as $\left\langle f \right\rangle  = \sum\nolimits_{\bf{n}} {f\left( {\bf{n}} \right)} Q\left( {{\bf{n}},t} \right)$ in general). In terms of the microscopic rates, we have
%$d \langle {{n_x}} \rangle /dt  = \sum\nolimits_y {[\langle {{n_y}W^\mu_{y \to x} {\left( {\bf{n}} \right)}} \rangle  - \langle {{n_x} W^\mu_{x \to y} {\left( {\bf{n}} \right)}} \rangle ]}$,
\begin{equation}\label{ddt_nx_nywyx_nxwxy}
\frac{d}{{dt}} \langle {{n_x}} \rangle  = \sum\limits_y {[\langle {{n_y}W^\mu_{y \to x} {\left( {\bf{n}} \right)}} \rangle  - \langle {{n_x} W^\mu_{x \to y} {\left( {\bf{n}} \right)}} \rangle ]}  \ ,
\end{equation}
which is so far an exact equation. Assuming the fluctuations around the mean value is small, {we can make the following approximation:}
%$\langle {n_x} \ W^\mu_{x \to y} {\left( {\bf{n}} \right)}\rangle  \approx \langle {n_x}\rangle \ W^\mu_{x \to y} {\left( \langle {\bf{n}} \rangle \right)}$.
\begin{equation}
\langle {n_x} \ W^\mu_{x \to y} {\left( {\bf{n}} \right)}\rangle  \approx \langle {n_x}\rangle \ W^\mu_{x \to y} {\left( \langle {\bf{n}} \rangle \right)} \ .
\end{equation}
Furthermore, since the players are homogeneous and the transition is taken randomly among the neighboring players, the microscopic transition rate $W_{x \to y}^\mu  ({\bf n}) = {w_{x \to y}} \ {n_y}$ is proportional to the number of players $n_y$ with strategy ${\cal S}_y$. Here ${w_{x \to y}}$ is the transition rate between any pair of players, taken for the mean-field configuration. Therefore, we have
\begin{equation}
\langle {n_x} W_{x \to y}^\mu \left( {\bf{n}} \right)\rangle  \approx \left\langle {{n_x}} \right\rangle {w_{x \to y}} \langle {{n_y}} \rangle
\end{equation}
under the mean field approximation.

Note that the averaged value $\langle {\bf{n}} \rangle $ is now taken as the input for the microscopic transition rate. {By defining ${\rho _x}\left( t \right) \equiv \left\langle {{n_x}} \right\rangle /N$, we have}
%$ d {\rho _x}\left( t \right) /dt = \sum\nolimits_y {\left[ {{\rho _y}\left( t \right){w_{y \to x}} \ {\rho _x}\left( t \right) - {\rho _x}\left( t \right){w_{x \to y}} \ {\rho _y}\left( t \right)} \right]}$,
\begin{equation}\label{dynamical_eom_mf}
\frac{d}{{dt}}{\rho _x}\left( t \right) = \sum\limits_y {\left[ {{\rho _y}\left( t \right){w_{y \to x}} \ {\rho _x}\left( t \right) - {\rho _x}\left( t \right){w_{x \to y}} \ {\rho _y}\left( t \right)} \right]} \ ,
\end{equation}
which is in exactly the form of the ALV equation (see Eq.~(\ref{ALV_eq})). The remaining task is to connect this equation with the game and the update rules, in which our
results can be applied effectively. For
illustrative purposes, we will present the results for the game called prisoner's dilemma.
%%%%%%%%%%%%%%%%%%%%%%%%%%%

\begin{figure*}[t]
    \centering
    \includegraphics[width=1.8\columnwidth]{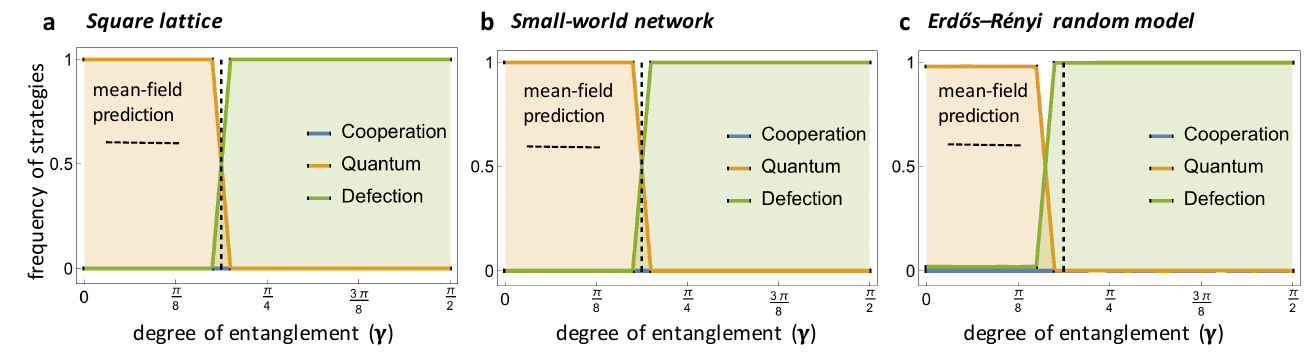}
    \caption{Numerical simulation on the anti-symmetric Lotka-Volterra (ALV) equation in Eq.~(\ref{dynamical_eom_mf}) for frequencies versus entanglement in different network topologies, namely (a) square lattice (grid of $100 \times 100$), (b) small-world network ($100 \times 100$ network with $p=0.01$), and (c) Erd\H{o}s-R\'{e}nyi model (with average degree $4$). Each data point is taken by playing the games after $10, 000$ steps.}
    \label{fig:topo}
\end{figure*}

\section{Microscopic interaction}
{In the model, the players are assumed to interact through a symmetric game, where an entangled quantum states are supplied as an additional resource; This generalization of classical games is referred to as a quantum game, which is a branch of quantum information science. In fact, many protocols in quantum information theory can be viewed as a realization of a game. For example, quantum crytography is a game involving two agents who aim to establish secure communication against eavesdroppers. Quantum cloning can be casted as a game played with the nature. The problem of quantum state discrimination can be viewed as a game to optimize the payoffs of two players, who are betting for the outcomes encoded in quantum states. Testing local hidden variable theory versus quantum theory can be formalized as a game.}
% check Benjamin, S. C. & Hayden, P. M. Multiplayer quantum games. Physical Review A 64, 030301 (2001).

{\begin{table}[t]
\caption{Payoff table of classical prisoner dilemma}
\centering
%\caption{My caption}
%\label{my-label}
\begin{tabular}{l|lr|lr}
\hline
\hline
\multicolumn{1}{l|}{} & \multicolumn{2}{l|}{\ {\bf C}ooperate \ } & \multicolumn{2}{l}{\ \ \ {\bf D}efect \ \ \ } \\ \hline
\multirow{2}{*}{ \ {\bf C}ooperate \ } &  & $R$ \ \ \ &  & $T$ \ \ \ \\
 & \ \ \ $R$ &  & \ \ \ $S$ &  \\ \hline
\multirow{2}{*}{ \ {\bf D}efect \ } &  & $S$ \ \ \ &  & $P$ \ \ \ \\
 & \ \ \ $T$ &  & \ \ \ $P$ &  \\
 \hline
 \hline
\end{tabular}
\label{tab:classical_PD}
\end{table}}
{Our model can be applied to any quantum game. To make our discuss concrete, we assume that the rewards of the players are given by the payoff table in the game of prisoner's dilemma, which is one of the most well-known two-player games and is adopted to explain how the emergence of cooperative or altruistic behavior can be resulted by selfish competitors~\cite{Szabo2007,alonso-Sanz2018}.}

{In the game, each agent (or player) is assumed to have a well-defined goals or preferences that can be quantified by a utility function, which represents the satisfaction of the agents obtained from an outcome of a game; in other words, the agents make decisions to optimize the utility function. An important milestone of game theory is the discovery of Nash equilibrium, which determines the best responses to other agents' actions and hence suggests guidelines for resolving dilemmas.}

Specifically, each play has two options, either defection ({\bf D}) or cooperation
({\bf C}), and they have to make the decision simultaneously without communication. If a mutual cooperation is made, both players will be rewarded by an amount $R$. For mutual defection, both of them will be punished with $P$. If one cooperates and the other defects, then the cooperator gains the lowest payoff $S$ and the traitor gains temptation $T$. {The payoff table of the symmetric game is summarized in table~\ref{tab:classical_PD}.}

The values of the payoffs are not necessarily fixed, but they are required to have the following order: $T  > R  > P  > S $.
%\begin{equation}
%   T \ > \ R \ > \ P \ > \  S \ .
%\end{equation}
If one of the players (called A) chooses to cooperate, the payoff of the other player (B) is higher if he/she chooses to defect (as $T>R$). The same is true even if A chooses to defect; B still gain more by choosing defection (as $P>S$). If both players are rational, they should both choose defection, i.e., $({\bf D},{\bf D})$ , which is the { \it Nash equilibirum}~\cite{neal2017,hilhorst2018}. However, this outcome does not give them the highest payoffs, which creates the dilemma.
%More concretely, we assume that the rewards of the players are given by the payoff table in the game of prisoner's dilemma. Specifically, each play has two options, either defection ({\bf D}) or cooperation
%({\bf C}), and they have to make the decision simultaneously without communication. If a mutual cooperation is made, both players will be rewarded by an amount $R$. For mutual defection, both of them will be punished with $P$. If one cooperates and the other defects, then the cooperator gains the lowest payoff $S$ and the traitor gains temptation $T$.
%
%The values of the payoffs are required to have the following order:
%%$T  > R  > P  > S $.
%\begin{equation}
%   T \ > \ R \ > \ P \ > \  S \ .
%\end{equation}
%If one of the players (called A) chooses to cooperate, the payoff of the other player (B) is higher if he/she chooses to defect (as $T>R$). The same is true even if A chooses to defect; B still gain more by choosing defection (as $P>S$). If both players are rational, they should both choose defection, i.e., $({\bf D},{\bf D})$, which is the { \it Nash equilibrium}~\cite{neal2017,hilhorst2018}. However, this outcome does not give them the highest average payoffs, which creates the dilemma.

\section{Making decisions with entanglement}
 Quantum game theory provides an extra resource for the
players in classical games, namely quantum entanglement ~\cite{Horodecki2009a},
which is a form of non-classical correlation that is essential for many tasks in quantum information processing~\cite{Xu2012,Zhang2012c}. In short, the quantum players can
adopt a mixed strategy guided by quantum probabilities. For
the game of prisoner's dilemma, the players can receive payoffs
no longer limited by the Nash equilibrium in the classical
game.

In the microscopic interactions of the network of prisoner's dilemma, the players are initially shared with an entangled state,
%$\left| {{\psi _{\rm ini}}} \right\rangle  = J\left| 0 \right\rangle  \otimes \left| 0 \right\rangle$,
{\begin{equation}
\left| {{\psi _{\rm ini}}} \right\rangle  = J\left| 0 \right\rangle  \otimes \left| 0 \right\rangle \ ,
\end{equation}}
where $J = \exp ( {i{\textstyle{\gamma  \over 2}}{\sigma _y} \otimes {\sigma _y}} )$ is an entangling operator (see Fig.~\ref{fig:prisoner} (a)). {The concurrence, which is regarded as a general measure of the entanglement of two-qubit state~\cite{Hill1997}, is $\sin \gamma$ for the initial state, where $\gamma$ is defined as the entanglement parameter.} The degree of entanglement of the initial state is strengthened by increasing the value of the real parameter $\gamma \in [0,\pi/2]$; when $\gamma =0$, no entanglement exists in the initial state, but the entanglement reaches its maximum when $\gamma = \pi/2$. However, the shared entanglement along does not allow the players to communicate. For example, in the protocol of quantum teleportation, a classical communication is needed at the end.

The players in the quantum game~\cite{Eisert1998,Li2014a} are allowed to apply, {\it independently}, a (single-qubit) unitary operation $U$ of the form,
{\begin{equation}
{U}(\theta,\phi) = \left( \begin{matrix} c & s \\  - s & c^* \end{matrix} \right) \ ,
\end{equation}}
where $c =  e^{i\phi}\cos\theta/2$ and $s= \sin\theta/2$,
%\begin{equation}
%{U}(\theta,\phi)=\left(
%                       \begin{array}{cc}
%                         e^{i\phi}\cos\theta/2 & \sin\theta/2\\
%                         -\sin\theta/2 & e^{-i\phi}\cos\theta/2 \\
%                       \end{array}
%                     \right) \ ,
%\end{equation}
on their own qubit. However, we restrict the available choices for the angles in a parameter sets, namely $(\theta ,\phi ) \in \left\{ {(0,0),(\pi ,0),(0,\pi /2)} \right\}$. These three choices correspond to three available strategies, ${\cal S} = \{ {\bf C} , {\bf D} , {\bf Q} \}$; they stand for $\bf C$ooperation, $\bf D$efection, and $\bf Q$uantum, given by
%$ {\bf C}  \Leftrightarrow  {U}(0,0)$, $ {\bf D}  \Leftrightarrow  {U}(\pi,0)$, and $ {\bf Q}  \Leftrightarrow  {U}(0,\pi/2)$.
{\begin{equation}\label{stra:CDQ}
 {\bf C}  \Leftrightarrow  {U}(0,0) \ , \quad
 {\bf D}  \Leftrightarrow  {U}(\pi,0) \ , \quad
 {\bf Q}  \Leftrightarrow  {U}(0,\pi/2) \ .
\end{equation}}
After that, the qubits from the players $A$ and $B$ are sent to a quantum machine that applies a unitary operator $J^\dagger$ to the qubits, which yields a final state,
%$\left| {{\psi _{{\text{fin}}}}} \right\rangle  = {J^\dag } ({U_A} \otimes {U_B}) \left| {{\psi _{{\text{ini}}}}} \right\rangle$.
{\begin{equation}
\left| {{\psi _{{\text{fin}}}}} \right\rangle  = {J^\dag } ({U_A} \otimes {U_B}) \left| {{\psi _{{\text{ini}}}}} \right\rangle  \ .
\end{equation}}
Finally, a joint quantum measurement, $\left\{ {{\Pi _{00}},{\Pi _{01}},{\Pi _{10}},{\Pi _{11}}} \right\}$, where ${\Pi _{ij}} \equiv \left| {ij} \right\rangle \left\langle {ij} \right|$ for $i,j \in \{ 0, 1\}$, is then applied to the final state to determine the expectation value of the individual utilities or payoffs $p_A$ and $p_B$ of the players, based on the payoff table of the classical game, i.e.,
\begin{equation}\label{eq:PI00011011}
p_{A} = p_B = R\left\langle {{\Pi _{00}}} \right\rangle  + S\left\langle {{\Pi _{01}}} \right\rangle  + T\left\langle {{\Pi _{10}}} \right\rangle  + P\left\langle {{\Pi _{11}}} \right\rangle \ ,
\end{equation}
%\begin{equation}
%  p_{A} = p_B = R\left\langle {{\Pi _{00}}} \right\rangle  + S\left\langle {{\Pi _{01}}} \right\rangle  + T\left\langle {{\Pi _{10}}} \right\rangle  + P\left\langle {{\Pi _{11}}} \right\rangle \ ,
%\end{equation}
where $\langle {{\Pi _{ij}}} \rangle  = {\left| {\left\langle {ij} \right.\left| {{\psi _{{\text{fin}}}}} \right\rangle } \right|^2}$, for $i,j \in \{ 0, 1\}$, represents the probability for the outcome $(i,j)$.

{The payoff of one of the players is listed in Fig.~\ref{fig:prisoner} (c). When one of the players adopts the quantum strategy~$\bf Q$, the other player's payoff is given by
%$\max \left\{ {R\lambda  + P(1 - \lambda ),\;T\lambda  + S(1 - \lambda ),\;R} \right\}$,
\begin{equation}
\max \left\{ {R\lambda  + P(1 - \lambda ),\;T\lambda  + S(1 - \lambda ),\;R} \right\} \ ,
\end{equation}
where $\lambda \equiv \cos^2\gamma $. When $\lambda  > {\textstyle{{R - S} \over {T - S}}}$, the maximum value $R$ is achievable by the other player who also adopts the quantum strategy $\bf Q$. In this regime, the choice of the quantum strategies $({\bf Q},{\bf Q})$ represents a new Nash equilibrium, resulting better payoffs for both players, as compared with the Nash equilibrium $({\bf D},{\bf D})$ associated with classical strategies.}

\section{Fixed-points of evolutionary network}
 We denote the population of the three strategies by a vector ${\bf n} = {\left[ {{n_C},{n_D},{n_Q}} \right]^T}$, where ${n_C} + {n_D} + {n_Q} = N$, and ${\rho _X}\left( t \right) \equiv \left\langle {{n_X}} \right\rangle /N$, {keeping tracks of the number of players~$n_X$ who has adopted a certain strategy ${\cal S}_X$ at each moment of time.} {In the mean-field approach, the transition rate is determined by Eq.~(\ref{Fermi_function}) for ${\cal S}_X,{\cal S}_Y \in \{ {\bf C, D ,Q} \}$, i.e.,
\begin{equation}\label{transition_prob_W}
  w({{\cal S}_X} \to {{\cal S}_Y}) = {\left( {1 + {e^{ - ({P_X} - {P_Y})/T}}} \right)^{ - 1}} \ ,
\end{equation}}
where the average payoffs $P_X, P_Y \in \{ P_C, P_D, P_Q\}$ are determined by the quantum game of prisoner's dilemma through the following relation:
\begin{equation}\label{Ps_relatedby_rhos}
\left[ {\begin{array}{*{20}{c}}
  {{P_C}} \\
  {{P_D}} \\
  {{P_Q}}
\end{array}} \right] =  \left[ {\begin{array}{*{20}{c}}
  R&S&{{R_\lambda}} \\
  T&P&{{T_\lambda }} \\
  {{R_\lambda}}&{{S_\lambda}}&R
\end{array}} \right]\left[ {\begin{array}{*{20}{c}}
  {{\rho _C}} \\
  {{\rho _D}} \\
  {{\rho _Q}}
\end{array}} \right] \ ,
\end{equation}
where $R_\lambda \equiv R \lambda + P(1-\lambda)$, $T_\lambda \equiv T\lambda + S (1- \lambda)$, and $S_\lambda \equiv S \lambda + T (1-\lambda)$ {with $\lambda=\cos^2\gamma$ inversely proportional to the degree of entanglement.}

Note that we can always write
\begin{equation}\label{hyper_tangents}
 w({{\cal S}_X} \to {{\cal S}_Y}) - w({{\cal S}_Y} \to {{\cal S}_X}) \equiv \tanh {\Delta_{XY}} \ ,
\end{equation}
where ${\Delta _{XY}} \equiv ({P_X} - {P_Y})/2T$ for $X,Y \in \{C,D,Q\}$. Consequently, the dynamical (mean-field) equation of motion (see Eq.~(\ref{dynamical_eom_mf})) can
now be written, in the matrix form, as follows,
\begin{equation}\label{MicroALV}
\begin{split}
&\frac{d}{dt}\left[ {\begin{array}{*{20}{c}}
  {{\rho_C}} \\
  {{\rho_D}} \\
  {{\rho_Q}}
\end{array}} \right] = \\
& \left[ {\begin{array}{*{20}{c}}
  0&\tanh {\Delta_{CD}}&\tanh {\Delta_{CQ}} \\
  \tanh {\Delta_{DC}}&0&\tanh {\Delta_{DQ}} \\
  \tanh {\Delta_{QC}}&\tanh {\Delta_{QD}}&0
\end{array}} \right]\left[ {\begin{array}{*{20}{c}}
  {{\rho _C}} \\
  {{\rho _D}} \\
  {{\rho _Q}}
\end{array}} \right] \ ,
\end{split}
\end{equation}
which is an ALV equation. To understand the long-time behavior of this set of equation, we look for the fix-point solution of the equations.

Based on the analysis presented previously, the fixed points are given by
\begin{equation}
{[\rho _C^\infty ,\rho _D^\infty ,\rho _Q^\infty ]^T} = \{ {[1,0,0]^T},{[0,1,0]^T},{[0,0,1]^T}\} \ .
\end{equation}
The stability of these fixed points are determined by the payoff table of the games {(see Fig.~\ref{fig:prisoner} (c)),} {i.e., it depends on entanglement parameter $\gamma$. A bifurcation for the evolutionary network will occur by changing the degree of of entanglement. Then the corresponding Jacobian matrix for the three fixed points takes the same form as the Eq.~(\ref{Fermi_function}), whose eigenvalues are also easy to be obtained.}
Recall that in general we have exactly one stable, one unstable, and one saddle fixed point.

{Let us now consider how the use of entanglement resources allows us to create a bifurcation for the evolutionary network. More specifically, we shall show how the increase of entanglement stabilize the fixed points at ${\left[ {0,0,1} \right]^T}$. To this end, the entanglement parameter has to exceed a critical value,}
%$\gamma  > \gamma_{*1} \equiv {\cos ^{ - 1}}\sqrt {(R - S)/(T - S)}$.
\begin{equation}\label{define_gamma*1}
  \gamma  > \gamma_{*1} \equiv {\cos ^{ - 1}}\sqrt {(R - S)/(T - S)} \ .
\end{equation}

{\it Proof:} To make ${[0,0,1]^T}$ a stable fixed point, {we have to make sure its Jacobian matrix contains two negative eigenvalues, which is equivalent to the conditions that ${P_Q} - {P_C} > 0$ and ${P_Q} - {P_D} > 0$,} {due to the fact that $\tanh x$ is positive (negative) when is positive (negative).} {From Eq.~(\ref{Ps_relatedby_rhos}), we have ${P_Q} - {P_D} = R - {T_\lambda } = R - T{\cos ^2}\gamma  - S{\sin ^2}\gamma$ and ${P_Q} - {P_C} = R - {R_\lambda } = R - R{\cos ^2}\gamma  - P{\sin ^2}\gamma $. Therefore, the conditions for the fixed point at ${\left[ {0,0,1} \right]^T}$ to be stable are,
  \begin{gather}
  T \ {\cos ^2}\gamma  + S \ {\sin ^2}\gamma  < R \ , \hfill \\
  R \ {\cos ^2}\gamma  + P \ {\sin ^2}\gamma  < R  \ . \hfill
\end{gather}}
{Recall that for the prisoner's dilemma game, $ T> R > P> S$. The second inequality is always satisfied for any value of~$\gamma$. The first inequality is not valid in the limit where $\gamma \to 0$, but it can be satisfied whenever $\gamma$ exceeds a critical value,
\begin{equation}\label{define_gamma*1}
  \gamma  > \gamma_{*1} \equiv {\cos ^{ - 1}}\sqrt {(R - S)/(T - S)} \ .
\end{equation}}
It makes that the point ${[0,0,1]^T}$ is the stable fixed one when $\gamma  > \gamma_{*1} $.
%The first condition is always satisfied, as ${P_Q} - {P_C} = R - {R_\lambda } = (R - P)  {\sin ^2}\gamma$ is always positive for the order  $ T> R > P> S$ in the game of prisoner's dilemma. The second condition yields the following inequality, $R-T{\cos ^2}\gamma-S{\sin ^2}\gamma> 0$, which implies ${\cos ^2}\gamma  < (R - S)/(T - S)$, i.e.
%$\gamma  > \gamma_{*1} \equiv {\cos ^{ - 1}}\sqrt {(R - S)/(T - S)}$.
%Similarly, to stabilize ${\left[ {0,1,0} \right]^T}$, it needs to satisfy $  S{\cos ^2}\gamma  + T{\sin ^2}\gamma  < P $, which means that ${\cos ^2}\gamma  > (T - P)/(T - S)$ or
%$\gamma  < \gamma_{*2} \equiv {\cos ^{ - 1}}\sqrt {(T - P)/(T - S)}$.
%\begin{equation}\label{define_gamma*2}
%\gamma  < \gamma_{*2} \equiv {\cos ^{ - 1}}\sqrt {(T - P)/(T - S)} \ .
%\end{equation}

On the other hand, for the fixed point at ${\left[ {0,1,0} \right]^T}$, we have ${P_Q} - {P_D} = {S_\lambda } - P = S{\cos ^2}\gamma  + T{\sin ^2}\gamma  - P$ and ${P_D} - {P_C} = P - S$. The conditions for being a stable fixed point are ${P_D} - {P_C} > 0$ and ${P_Q} - {P_D} < 0$. The first one implies that $P>S$, which is always satisfied. The second one implies that
\begin{equation}
  S{\cos ^2}\gamma  + T{\sin ^2}\gamma  < P \ ,
\end{equation}
which means that ${\cos ^2}\gamma  > (T - P)/(T - S)$ or
\begin{equation}\label{define_gamma*2}
\gamma  < \gamma_{*2} \equiv {\cos ^{ - 1}}\sqrt {(T - P)/(T - S)} \ .
\end{equation}
We now have two critical values, namely $\gamma_{*1}$ and $\gamma_{*2}$, as defined in Eq.~(\ref{define_gamma*1}) and (\ref{define_gamma*2}). The former stabilizes the fixed point at ${\left[ {0,0,1} \right]^T}$ whenever $\gamma > \gamma_{*1}$, and the latter stabilizes ${\left[ {0,1,0} \right]^T}$ whenever $\gamma < \gamma_{*2}$. {If the values of the payoffs are chosen as
\begin{equation}
  T - P = R - S \ ,
\end{equation}
then it makes the two critical values of the above fixed points coincide with each other, i.e., ${\gamma _{*1}} = {\gamma _{*2}}$.}
As a result, by adjusting the entanglement parameter $\gamma$, the behavior of the evolutionary network can be changed drastically across the critical point. In the case, a bifurcation occurs that the equilibrium state changes abruptly from the one dominated by the  defection strategy ({\bf D})  to the one dominated by the quantum strategy ({\bf Q}).

To realize such a bifurcation, the payoffs of the game can be assigned as follows: $T = 1 + r$, $R=1$, $P=0$, and $S=-r$, for any $r>0$. Consequently, the phase boundary is given by the following relation:
\begin{equation}\label{eq:phase_boundary}
r_* = (1 - {\cos ^2}\gamma_* )/(2{\cos ^2}\gamma_*  - 1) \ ,
\end{equation}
which is represented by the white dotted line in Fig.~\ref{fig:phase}. Our numerical simulation of the ALV equation in Eq.~(\ref{dynamical_eom_mf}) reveals that similar critical transition exists for different network topologies (see Fig.~\ref{fig:topo}), where  the bifurcation points are in good agreement with the mean-field prediction in Eq.~(\ref{eq:phase_boundary}). This completes our analysis on the bifurcation mechanism.

For completeness, let us look at the remaining fixed point at $[1,0,0]^T$. The relevant quantities are ${P_D} - {P_C} = T - R$ and ${P_Q} - {P_C} = {R_\lambda} - R = R{\cos ^2}\gamma  + P{\sin ^2}\gamma  - R$. Since $T>R$, it is always true that $P_D \ge P_C$. Consequently, the fixed point $[1,0,0]^T$ can never become a stable one. To investigate further, since $R>P$, it is also true that $R \ {\cos ^2} \gamma  + P \ {\sin ^2} \gamma  - R < 0$, which implies that $P_C > P_Q$. Therefore, its Jacobian matrix contains exactly one positive and one negative eigenvalues; the fixed point at $[1,0,0]^T$ is always a saddle fixed point.

\section{Conclusion}
In summary, we presented a novel bifurcation mechanism for non-linear dynamical systems governed by the ALV equation. The bifurcation can be observed whenever the stable and unstable fixed points are exchanged by a control parameter. As an application, we focused on evolutionary networks, where the control parameter is characterized by the shared entanglement in the microscopic interaction. Furthermore, we performed numerical simulations to verify that such a bifurcation phenomenon is robust against the change in the network topology. In principle, the bifurcation mechanism can occur in other dynamical systems governed by the ALV equations as well, including predatory-prey oscillations in population biology, condensation of bosons far from equilibrium, plasma oscillation, kinetics of chemical reactions, where the dynamics is governed by the ALV equation.
\section*{Acknowledgements}
X.Y. and A.L. thank the support by the Grand Project
``Network Algorithms and Digital Information'' of the Institute of Software, Chinese Academy of Sciences, by an NSFC grant No. 61161130530 and No. 11405093, by a High-Tech Program (863) Grant No. 2012AA8113011, and by a China Basic Science Program (973) Grant No. 2014CB340302, M.-H.Y. and X.-K.S acknowledges support by the National Natural Science Foundation of China (11875160),  the NSFC-Guangdong Joint Fund (U1801661), the Guangdong Innovative and Entrepreneurial Research Team Program (No.~2016ZT06D348),  Natural Science Foundation of Guangdong Province (2017B030308003), the Science, Technology and Innovation Commission of Shenzhen Municipality (JCYJ20170412152620376, JCYJ20170817105046702, ZDSYS201703031659262), and the Postdoctoral Science Foundation of China (No.2018M632195).

\end{document}